\documentclass[pre,onecolumn]{revtex4}
\usepackage{amssymb}
\usepackage{amsmath}
\usepackage{graphicx}
\begin{document}
\title{Nonequilibrium dynamics of coupled oscillators under the shear-velocity boundary condition}
\author{Hidetsugu Sakaguchi}
\address{Interdisciplinary Graduate School of Engineering Sciences, Kyushu University,  Kasuga, Fukuoka 816-8580, Japan}
\begin{abstract}
Deterministic and stochastic coupled oscillators with inertia are studied on the rectangular lattice under the shear-velocity boundary condition. Our coupled oscillator model exhibits various nontrivial phenomena and there are various relationships with wide research areas such as the coupled limit-cycle oscillators, the dislocation theory, a block-spring model of earthquakes, and the nonequilibrium molecular dynamics. We show numerically several unique nonequilibrium properties of the coupled oscillators.  We find that the spatial profiles of the average value and variance of the velocity become non-uniform when the dissipation rate is large. The probability distribution of the velocity sometimes deviates from the Gaussian distribution. The time evolution of kinetic energy becomes intermittent when the shear rate is small and the temperature is small but not zero. The intermittent jumps of the kinetic energy cause a long tail in the velocity distribution.\end{abstract}
\maketitle
\section{Introduction}
Coupled phase oscillators called the Kuramoto model have been intensively studied by many authors as a simple solvable model of collective synchronization~\cite{Kuramoto, Bonilla, Strogatz}.  There are many generalized models for the Kuramoto model. One model includes the inertia term or the second derivative of the phase variable~\cite{Tanaka, Sakaguchi}. The coupled phase oscillators on square or cubic lattices was called oscillator lattices and the phase transition via the collective synchronization was studied in the finite-dimensional systems~\cite{Sakaguchi2,Hong}.
 
In the previous paper, we demonstrated that the vortex motion in the oscillator lattices is closely related to the dislocation motion in solids~\cite{Sakaguchi3}. The dislocation motion is important to understand the mechanical properties of the solid under the shear stress~\cite{Read, Cottrell}. 
The correspondence between the coupled phase oscillators and lattice dynamics is as follows. The phase variable in the oscillator lattice with inertia is interpreted as the one-dimensional displacement in the $z$ direction in a three-dimensional crystal, and the displacement is assumed to be uniform in the $z$ direction. That is,  the displacement $z_{i,j,k}$ at the $(i,j,k)$ site takes the same value for all $k$'s, and is expressed as $z_{i,j}$. The vortex in the oscillator lattice corresponds to the screw dislocation in the crystal. The shear stress on the boundaries corresponds to the external force at the boundaries of the oscillator lattices. We found that the vortex begins to move if the shear stress is beyond a critical force, which corresponds to the Peierls force in the dislocation theory. The spontaneous generation of vortices and the complex motion with nonzero frequencies correspond to the slip motion or the plastic flow in solids. We further studied the oscillator lattices under external noises, and found that the vortex motion occurs under the critical force owing to the fluctuation effect, which corresponds to the dislocation motion at a finite temperature~\cite{Sakaguchi4}. 

The coupled oscillators under the shear boundary conditions are also related to the mechanics of earthquakes and faulting~\cite{Scholz}. The block-spring models of earthquakes have been studied by many authors~\cite{Langer, Sakaguchi5}. In the block-spring model, each block has a mass and is coupled with the neighboring blocks through a linear spring. The shear force is applied from the top plate moving with a small velocity. A large slip motion in the block-spring model corresponds to an earthquake.

The deterministic chaos and fluctuation in the oscillator lattice under the shear stress is related to the nonequilibrium statistical mechanics. The nonequilibrium statistical mechanics has been numerically studied with the molecular dynamics by many authors~\cite{Evans, Ashurst}. Various nonequilibrium properties such as the shear viscosity~\cite{Liem} and heat conductivity~\cite{Lepri} have been investigated by numerical simulations of Newton's equation of motion of many particles. The relationship between the reversible equations and irreversible behavior was discussed by several authors.~\cite{Holian}. 
In the molecular dynamics of fluids, interacting pairs of molecules change with time. In our oscillator lattice models, each oscillator interacts only with the nearest neighbors, but a large displacement occurs by the phase slip between the neighboring oscillators. In this paper, we will discuss several nonequilibrium properties and velocity distributions in the coupled oscillators under the shear velocity boundary conditions. 

\section{Deterministic coupled oscillators under the shear-velocity boundary condition}
In this section, we consider deterministic coupled oscillators with inertia on the rectangular lattice:  
\begin{equation}
\frac{d^2 z_{i,j}}{dt^2}=K_x\sum_{i^{\prime}=i-1,i+1}(z_{i^{\prime},j}-z_{i,j})+K_y\sum_{j^{\prime}=j-1,j+1}\sin (z_{i,j^{\prime}}-z_{i,j})-d\frac{dz_{i,j}}{dt},
\end{equation}
where $z$ is the displacement in the $z$ direction  at the lattice point of $(i,j)$,  $d$ is the coefficient of resistance force in proportion to the velocity, and $(i^{\prime},j^{\prime})$'s are the nearest neighbor sites of the $(i,j)$ site on the rectangular lattice of $L_x\times L_y$. The linear and sinusoidal couplings are assumed respectively in the $x$- and $y$-directions. The coupling strength $K_x$ and $K_y$ are set to be 1 in this paper. 
Periodic boundary conditions are imposed at $i=1$ and $i=L_x$. In the $y$ direction, the boundary conditions corresponding to the shear velocity are imposed, that is, $z_{i,j}=0$ at $j=0$ and $z_{i,j}=v_0 t$ at $j=L_y+1$. 
The numerical simulation was performed with the Runge-Kutta method with timestep $\Delta t=0.005$. For $d=0$ and $v_0=0$, the coupled oscillator system has the Hamiltonian:
\begin{equation}
H=\sum_{i,j}\frac{1}{2}\left (\frac{dz_{i,j}}{dt}\right )^2+\frac{1}{2}\sum_{i,j}\sum_{i^{\prime}=i-1,i+1}(z_{i^{\prime},j}-z_{i,j})^2+\sum_{i,j}\sum_{j^{\prime}=j-1,j+1}\{1-\cos(z_{i,j^{\prime}}-z_{i,j})\}.
\end{equation}

For $L_x=L_y=1$, a single oscillator obeys the equation
\begin{eqnarray}
\frac{dz}{dt}&=& v,\nonumber\\
\frac{dv}{dt}&=&\sin(v_0t-z)-\sin z-dv.
\end{eqnarray}
Equation (3) is a conservative (dissipative) system at $d=0$ ($d\ne 0$) since the phase space volume decays as $e^{-dt}$. Figure 1(a) shows the Poincar\'e map (stroboscopic mapping) in the phase space of $({\rm mod}(z,2\pi),v)$ at $t=2\pi n/v_0$ ($n$: integer) for $v_0=0.1$ and $d=0$. The initial condition is $z(0)=0$ and $v(0)=0.05001$. A chaotic dynamics is observed owing to the periodic forcing term $\sin(v_0t-z)$. Figure 1(b) shows the velocity distribution of $v$. The average velocity  is 0.05. The velocity is confined between $-2.78$ and 2.88. 
\begin{figure}[h]
\begin{center}
\includegraphics[height=4.5cm]{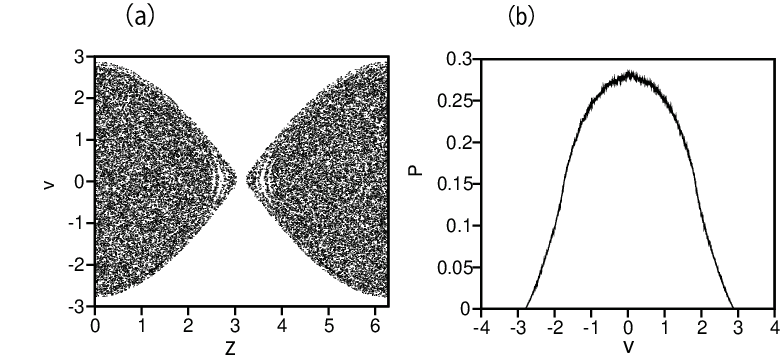}
\end{center}
\caption{(a) Poincar\'e map in the phase space of $({\rm mod}(z,2\pi),v)$ at $t=2\pi n/v_0$ for  $v_0=0.1$ and $d=0$. (b) Velocity distribution $P(v)$ for $v_0=0.1$ and $d=0$.}
\label{fig1}
\end{figure}
\begin{figure}[h]
\begin{center}
\includegraphics[height=4.5cm]{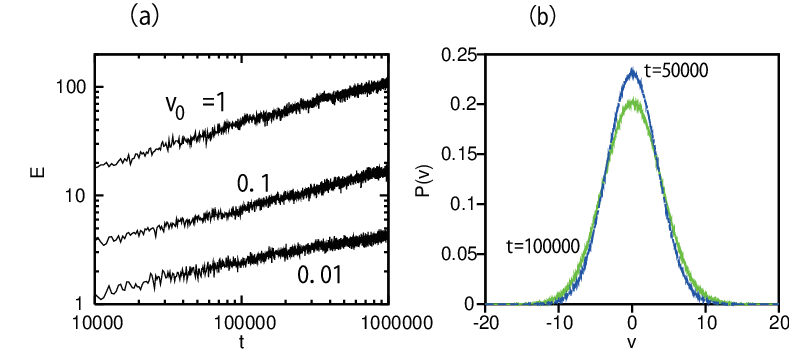}
\end{center}
\caption{(a) Time evolutions of the kinetic energy at $v_0=0.01$, 0.1, and 1 for $d=0$. (b) Probability distributions $P(v)$ at $v_0=0.1$ near $t=50000$ (blue line)  and $100000$ (green line).}
\label{fig2}
\end{figure}
\begin{figure}[h]
\begin{center}
\includegraphics[height=4.cm]{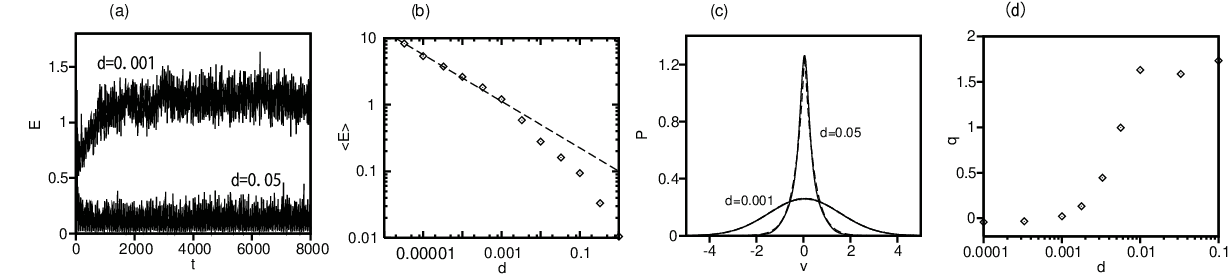}
\end{center}
\caption{(a) Time evolutions of the kinetic energy at $v_0=0.1$ for $d=0.001$ and 0.05. (b) Average kinetic energy $\langle E\rangle$ at $v_0=0.1$ as a function of $d$. The dashed line is $\langle E\rangle\propto t^{-\alpha}$ with $\alpha=0.35$.  (c) Probability distributions $P(v)$ at $d=0.05$ and $0.001$ for $v_0=0.1$, (d) Kurtosis of the probability distributions at several $d$'s. }
\label{fig1}
\end{figure}

For $L_x=300$ and $L_y=1$, the one-dimensional coupled oscillators obey 
\begin{eqnarray}
\frac{dz_i}{dt}&=& v_i,\nonumber\\
\frac{dv_i}{dt}&=&K\{\sin(v_0t-z_i)-\sin z_i\}-dv_i+K(z_{i+1}-2z_i+z_{i-1}).
\end{eqnarray}
Figure 2(a) shows the time evolution of the average value of the kinetic energy: $E=(1/L_x)\sum_{i=1}^{L_x}(1/2)v_i^2$ at $v_0=0.01$, 0.1, and 1 for $d=0$. The kinetic energy increases with time. The time evolution can be approximated as a power law $E\propto t^{\alpha}$ with $\alpha\simeq 0.27$, 0.35, and 0.41 respectively for $v_0=0.01$, 0.1, and 1. The increase of kinetic energy or heating occurs under the shear-velocity boundary conditions. The heating is not observed at $L_x=1$ as shown in Fig.~1. We have checked that the heating occurs for $L_x\ge 4$ at $v_0=0.1$. The critical size for the heating depends on $v_0$, that is, the critical size $L_{xc}$ was evaluated as 3, 8, and 13 respectively for $v_0=1$, $0.05$, and 0.03. The critical size decreases as $v_0$ is larger. 
Although the heating occurs due to the nonequilibrium boundary conditions, the mechanism is not clear. Figure 2(b) shows the probability distributions of $v$ at $v_0=0.1$ near $t=50000$ and $100000$, The probability distributions are well approximated at the Gaussian distribution, and the variance increases with time. The Gaussian distribution of the velocity corresponds to the Maxwell distribution of the ideal gas in the statistical mechanics. 

Figure 3(a) shows the time evolution of $E$ at $v_0=0.1$ and $d=0.001$ and 0.05. The increase of the kinetic energy stops owing to the dissipation term $-dv_i$  with $d=0.001$. 
The kinetic energy is fluctuating at $d=0.001$ and 0.05. The average value of $E$ and the fluctuation amplitude is the same order at $d=0.05$.  
Figure 3(b) shows the relationship between $d$ and the average kinetic energy $\langle E\rangle$ at $v_0=0.1$. The dashed line is $\langle E\rangle\propto t^{-\alpha}$ with $\alpha=0.35$. The numerical result of the time evolution of $E$ as $E\propto t^{\alpha}$ at $d=0$ suggests that $E$ satisfies an approximate equation $dE/dt=\beta E^{1-1/\alpha}$ at $d=0$.   
By adding the dissipation term of $-2dE$, the time evolution of the kinetic energy $E$ is assumed to be
\[\frac{dE}{dt}=\beta E^{1-1/\alpha}-2dE.\]
The stationary value of $E$ is expressed with 
\[E\propto d^{-\alpha}.\] 
Figure 3(b) shows that the power law $\langle E\rangle\propto d^{-\alpha}$ is satisfied for $d<0.001$. $\langle E\rangle$ deviates from the power law for $d>0.001$. 
Figure 3(c) shows the probability distributions $P(v)$ at $d=0.001$ and 0.05. 
The dashed lines are $P(v)=1/\sqrt{2\pi\sigma^2} \exp\{-(v-v_0/2)^2/(2\sigma^2)\}$ with $\sigma^2=2.35$ and $P(v)=1/(2\gamma)\exp(-|v-v_0/2|/\gamma)$ with $\gamma=0.39$. (The dashed line  $P(v)=1/\sqrt{2\pi\sigma^2} \exp\{-(v-v_0/2)^2/(2\sigma^2)\}$ is hardly seen because of the overlapping with the numerically obtained $P(v)$.) For small $d$, $P(v)$ is well approximated at the Gaussian distribution, however, $P(v)$ deviates from the Gaussian distribution and closer to the exponential distribution. Figure 3(d) shows the numerically obtained kurtosis $q=\langle(v-v_0)^4\rangle/(\langle (v-v_0)^2\rangle)^2-3$ as a function of $d$. The kurtosis is 0 for the Gaussian distribution and $q$ is a quantity expressing the deviation from the Gaussian distribution. Figure 3(d) shows that the deviation from the Gaussian distribution appears for $d>0.002$. 
The deviation of $P(v)$ from the Gaussian distribution at $d=0.05$ might be due to the large fluctuation of the kinetic energy compared to the average value, although the mechanism is not well understood. 

\begin{figure}[h]
\begin{center}
\includegraphics[height=4.5cm]{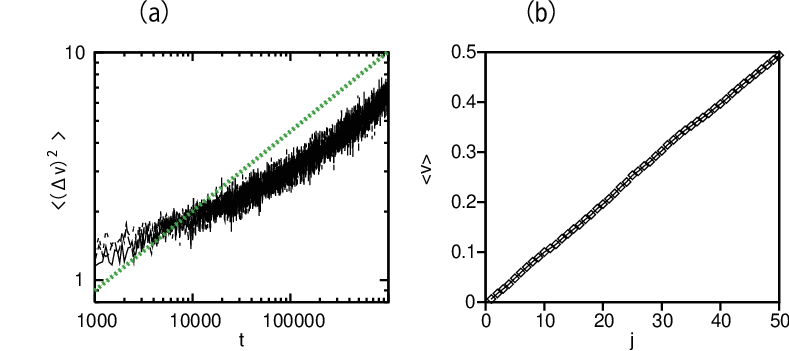}
\end{center}
\caption{(a) Time evolutions of the variance of $v_j$ for $j=5,30$, and $45$ at $d=0$ and $v_0=0.5$. The three lines overlap well except for small $t$. 
The dashed green line denotes the power law of $t^{\alpha}$ with 0.35. (b) Average velocity $\langle v_j\rangle$ at $d=0$ and $v_0=0.5$.}
\label{fig4}
\end{figure}
\begin{figure}[h]
\begin{center}
\includegraphics[height=4.5cm]{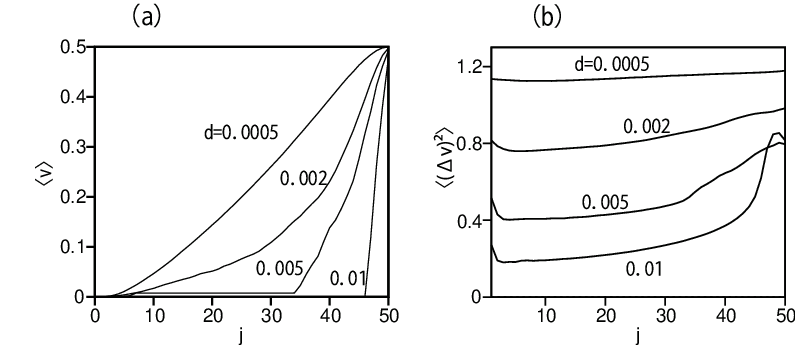}
\end{center}
\caption{(a) Average velocity $\langle v_j\rangle$ at $d=0.0005$, 0.002, 0.005, and 0.01 for $v_0=0.5$. (b) Variance $\langle (\Delta v_j)^2\rangle$ of the velocity at $d=0.0002$, 0.002 and 0.01. }
\label{fig5}
\end{figure}
\begin{figure}[h]
\begin{center}
\includegraphics[height=4.5cm]{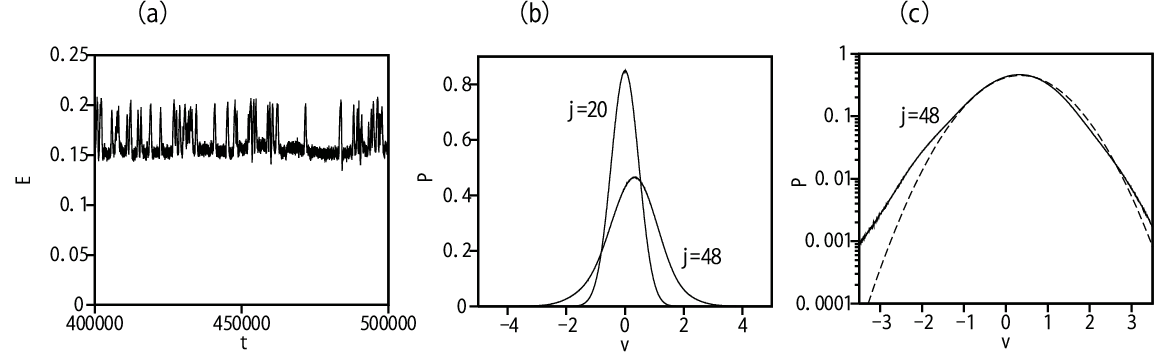}
\end{center}
\caption{(a) Time evolution of the kinetic energy per one oscillator: $E=\sum (1/2)v_{i,j}^2/(L_xL_y)$  at $d=0.01$ and $v_0=0.5$. (b) Probability distribution $P(v)$ of the velocity profiles at $j=20$ and $j=48$ for $d=0.01$. (c) Probability distribution $P(v)$ of the velocity profile in the semi-logarithmic scale at $j=48$ for $d=0.01$. The dashed line is $1/\sqrt{2\pi\sigma^2}\exp\{-(v-0.35)^2/(2\sigma^2)\}$ with $\sigma^2=0.785$. }
\label{fig6}
\end{figure}
Next, we show numerical results for a rectangular system of $L_x\times L_y=300\times 50$.
Figure 4(a) shows the time evolutions of the variance of $v_j$ for $i=1,2,\cdots,L_x$ at $j=5,30$, and $45$ for $d=0$ and $v_0=0.5$. The variance increases with time and do not depend on $j$. The dashed line denotes the power law of $t^{\alpha}$ with 0.35. The variance increases faster than a power law, but it is confirmed that the heating occurs also in this rectangular system. Figure 4(b) shows the profile of the average velocity $\langle v_j\rangle$ at $d=0$ and $v_0=0.5$. The average velocity changes linearly in $j$. The linear shear flow is characteristic in the Newton fluid with the linear viscosity. 

Figure 5(a) shows the profiles of the average velocity $\langle v_j\rangle$ at $d=0.0005$, 0.002, 0.005, and 0.01 for $v_0=0.5$. The average velocity changes almost linearly in $j$ at $d=0.005$ and the difference $\langle v_{j+1}\rangle-\langle v_j\rangle$ or the shear rate between the $j$th and $(j+1)$th layers is nearly constant.  However, the shear rate increases with $j$ at $d=0.002$. The shear rate takes a small value for $j<33$ and takes a large vale for $j>33$ at $d=0.005$. The average velocity changes rapidly from 0 to $v_0=0.5$ near $j=L_y$ at $d=0.01$. That is, the shear rate is almost zero for $j<46$ and takes a very large value for $j>46$ at $d=0.01$.  The localization of shear rate  near the boundary $j=L_y$ is a phenomenon similar to the one observed in the coupled noisy oscillators without inertia found in the previous paper~\cite{Sakaguchi4}. We showed that the shear localization occurs when $v_0$ is relatively large, and the vortex density is  high in the boundary region where the shear rate of the average velocity takes a large value and the vortex density takes a lower constant value in the bulk region where the average velocity is almost zero. The phenomena of shear or strain  localization are observed in various complex fluids and plastic materials such as the plug flow in the Bingham fluid~\cite{Bingham}, the shear banding~\cite{Salmon}, and the formation of shear zones in rocks~\cite{Rutter}, although the mechanism might be different. The reason of the shear localization in our model is not well understood yet. 
Figure 5(b) shows the profiles of the variance of the velocity $v_j$ at $d=0.0005$, 0.002, 0.005, and 0.01. The variance of the velocity corresponds to the temperature. The temperature is almost uniform for $d=0.0005$, but the temperature is higher near $j=L_y$ for $d=0.01$. This is the excess heating due to the shear of the velocity profile near $j=L_y$. Even if the average velocity is zero, the variance of the velocity takes a nonzero value and changes slowly for $j<46$ at $d=0.01$, that is, the oscillators are not stationary. 

Figure 6(a) shows the time evolution of the kinetic energy per each oscillator $E=(1/L_xL_y)\sum (1/2)v_{i,j}^2$  at $d=0.01$. Intermittently, $E$ takes a large value, which is related to the excess heating. Figure 6(b) shows the probability distribution of the velocity $v_j$ at $j=20$ and $j=48$ for $d=0.01$. The velocity distribution at $j=20$ is well approximated at the Gaussian distribution $\exp\{-v^2/(2\sigma^2)\}/\sqrt{2\pi\sigma^2}$ with $\sigma^2=0.22$. Figure 6(c) is the velocity distribution $P(v)$ in the semi-logarithmic scale at $j=48$ for $d=0.01$. The dashed line is $1/\sqrt{2\pi\sigma^2}\exp\{-(v-0.35)^2/(2\sigma^2)\}$ with $\sigma^2=0.785$. The velocity distribution at $j=48$ is not mirror symmetric around the peak value and deviated from the Gaussian distribution probably due to the shear velocity. 
\section{Coupled noisy oscillators under the shear-velocity boundary condition}
The coupled oscillator model at a finite temperature is assumed to be  
\begin{equation}
\frac{d^2 z_{i,j}}{dt^2}=K_x\sum_{i^{\prime}=i+1,i-1}(z_{i^{\prime},j}-z_{i,j})+K_y\sum_{j^{\prime}=j+1,j-1}\sin (z_{i,j^{\prime}}-z_{i,j})-d\frac{dz_{i,j}}{dt}+\xi_{i,j}(t),
\end{equation}
where $\xi_{i,j}(t)$ is the Gaussian white noise satisfying $\langle \xi_{i,j}(t)\xi_{i^{\prime},j^{\prime}}(t^{\prime})\rangle=2dT\delta_{i,i^{\prime}}\delta_{j,j^{\prime}}\delta(t-t^{\prime})$. Here, $T$ is a quantity corresponding to the temperature. The coupling constants are assumed to be $K_x=K_y=1$. If the periodic boundary conditions are imposed for all the four boundaries: $i=1$, $i=L_x$, $j=1$, and $j=L_y$, the thermal equilibrium distribution is realized. The probability distribution of the velocity $v_{i,j}$ is the Gaussian distribution
\[P\propto e^{-H/T} \propto e^{-\sum_{i,j}v_{i,j}^2/(2T)},\]
where $H$ is the total energy of Eq.~(2). 
The shear-velocity boundary conditions induce a nonequilibrium state.
Figure 7(a) shows the average velocity $\langle v_j\rangle$ as a function of $j$ at $T=2$, $d=0.01$ and $v_0=0.5$. By the effect of finite temperature, the velocity profile is more delocalized than the one shown in Fig.~5(a). The dashed blue line is an approximate line $0.5\exp\{-0.185(50-j)\}$ for the velocity profile. Figure 7(b) shows the variance profile of the velocity $v_j$. 
The variance is almost $T=2$ for $j<35$, which is almost equal to the strength of the noise, however, the variance is slightly larger than 2 near the boundary $j=51$. Figure 7(c) is the probability distributions at $j=20$ and 48. The probability distribution can be well approximated at the Gaussian distribution in this finite-temperature system. When $T$ is large, the stochastic motion is dominant as shown in Fig.~7. 

The length scale of the shear localization is evaluated by a quantity $\lambda=\sum_{j=1}^{L_y}(L_y+1-j)\langle v_j\rangle /\sum_{j=1}^{L_y}\langle v_j\rangle$.
Figure 8(a) shows $\lambda$ for several $d$'s at $T=2$ and $v_0=0.5$, and Fig.~8(b) shows $\lambda$ for several $v_0$'s at $T=0.5$ and $d=0.01$. The length scale $\lambda$ decreases with $d$ and $v_0$.
\begin{figure}[h]
\begin{center}
\includegraphics[height=4.5cm]{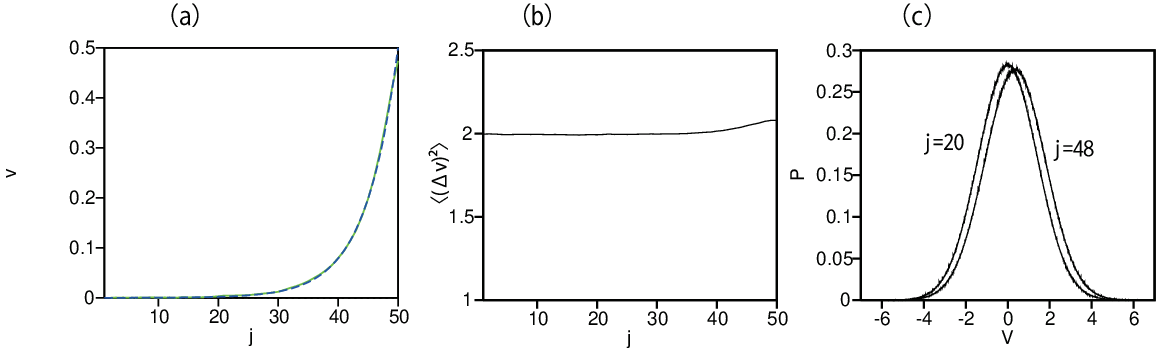}
\end{center}
\caption{(a) Profile of the average velocity $\langle v_j\rangle$ at $d=0.01$, $T=2$, and $v_0=0.5$. (b) Profile of the variance of the velocity $v_j$. (c) Probability distributions at $j=20$ and 48.}
\label{fig7}
\end{figure}
\begin{figure}
\begin{center}
\includegraphics[height=4.5cm]{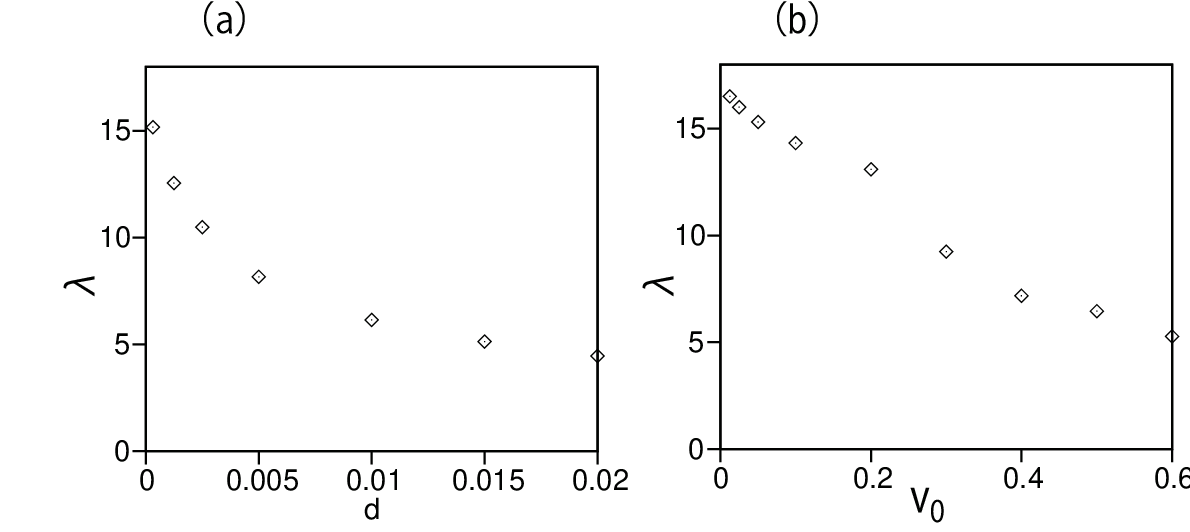}
\end{center}
\caption{(a) Length scale $\lambda=\sum_{j=1}^{L_y}(L_y+1-j)\langle v_j\rangle /\sum_{j=1}^{L_y}\langle v_j\rangle$ for several $d$'s at $T=2$ and $v_0=0.5$. (b) Length scale $\lambda$ for several $v_0$'s at $T=0.5$ and $d=0.01$.}
\label{fig5}
\end{figure}
\begin{figure}[h]
\begin{center}
\includegraphics[height=4.5cm]{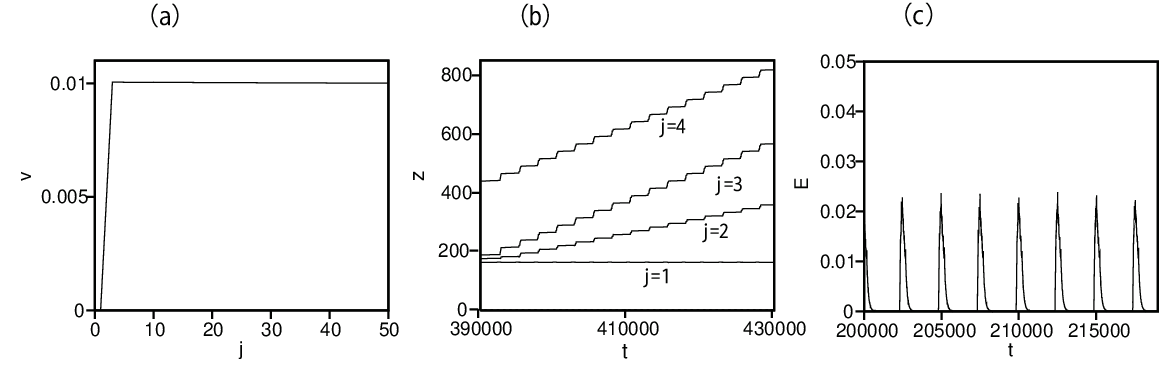}
\end{center}
\caption{(a) Profile of the average velocity $\langle v_j\rangle$ at $d=0.01$, $T=0$, and $v_0=0.01$. (b) Time evolutions of $z_{i,j}$ between $t=390000$ and 430000 at $(i,j)=(n/2,1)$, $(n/2,2)$, $(n/2,3)$, and $(n/2,4)$ for $d=0.01$, $T=0$ and $v_0=0.01$. (c) Time evolution of the kinetic energy $E$.}
\label{fig8}
\end{figure}
\begin{figure}[h]
\begin{center}
\includegraphics[height=4.cm]{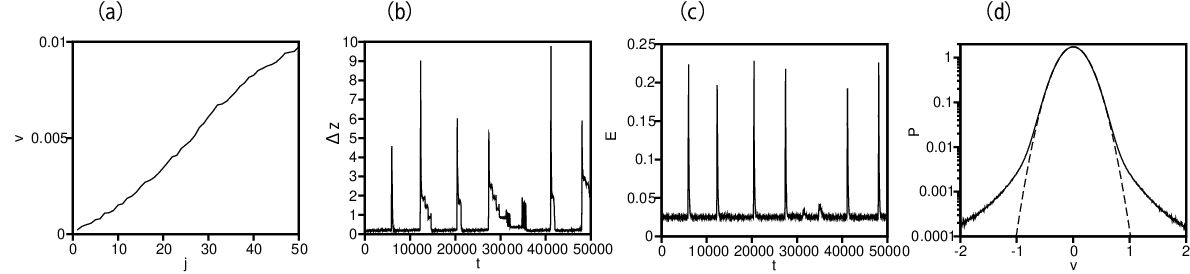}
\end{center}
\caption{(a) Profile of the average velocity $\langle v_j\rangle$ at $d=0.01$, $T=0.05$ and $v_0=0.01$. (b) Time evolution of the standard deviation $\Delta z$. (c) Time evolution of the kinetic energy $E$. (d) Semi-logarithmic plot of the probability distribution $P(v_j)$ at $j=20$ The dashed line is $\sqrt{2\pi \sigma^2}\exp\{-(v-0.00345)^2/(2\sigma^2)\}$ with $\sigma=0.05025$.}
\label{fig9}
\end{figure}
\begin{figure}[h]
\begin{center}
\includegraphics[height=4.5cm]{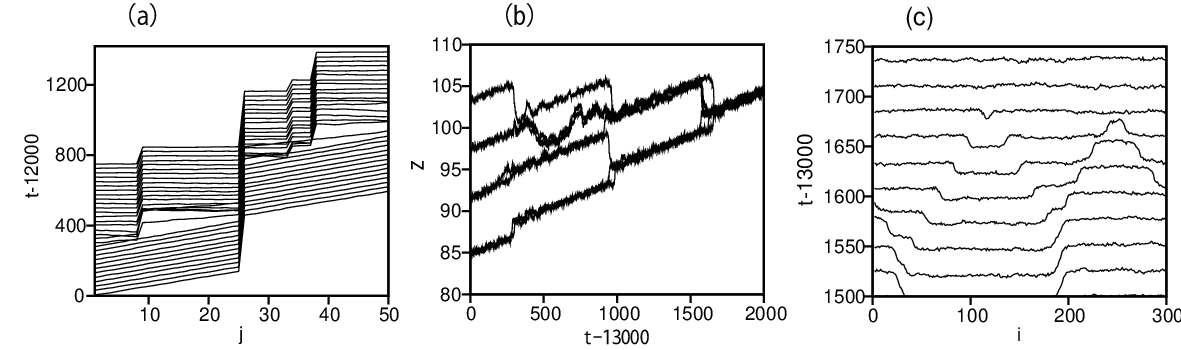}
\end{center}
\caption{(a) Time evolution of the prifile $z_{i,j}$ at $i=25$ from $t=12000$ to 12750. (b) Time evolutions of $z_{i,j}$ of six points $(i,j)=(1,36)$, $(51,36)$, $(101,36)$, $(151,36)$, $(201,36)$, and $(300,36)$ from $t=13000$ to 15000. (c) Time evolution of the profile $z_{i,j}$ at $j=36$ from $t=14500$ to 14750.}
\label{fig10}
\end{figure}

When $v_0$ and $T$ are small, a more regular motion appears. Figure 9(a) shows the profile of the average velocity $\langle v_j\rangle$ at $d=0.01$, $T=0$ and $v_0=0.01$. Figure 9(b) shows the time evolutions of $z_{i,j}$ between $t=390000$ and 430000 at $(i,j)=(n/2,1)$, $(n/2,2)$, $(n/2,3)$, and $(n/2,4)$ for $d=0.01$, $T=0$ and $v_0=0.01$. $z_{i,j}$ increases stepwise except for $j=1$. That is, the phase slip of $2n\pi$ occurs almost simultaneously.  
The period of the stepwise increase is around 2500, and the phase slip of $4\pi$ occurs at $j=2$ and the phase slip of $8\pi$ occurs for $j\ge 3$. The average velocity is zero at $j=1$ and takes the same value for $j\ge 3$, and 
the velocity profile has a jump from 0 to 0.01 at $j=2$ as shown in Fig.~9(a).  Figure 9(c) shows the time evolution of the kinetic energy $E$ for $T=0$, $v_0=0.01$, and $d=0.01$. The kinetic energy increases periodically due to the periodic phase slips, whose period is around 2500.  

Figure 10(a) shows the profile of the average velocity $\langle v_j\rangle$ at $d=0.01$, $T=0.05$ and $v_0=0.01$. The average velocity changes continuously with $j$ owing to the finite temperature effect and the profile might be approximated at the linear one. 
Figure 10(b) shows the time evolution of $\Delta z=\sqrt{\sum_j\sum_i(z_{i,j}-\sum_i z_{i,j}/L_x)^2/(L_xL_y)}$ at $d=0.01$, $T=0.05$ and $v_0=0.01$. Here, $\sum_i(z_{i,j}-\sum_i z_{i,j}/L_x)^2/L_x$ is the variance $\langle (\Delta z_{i,j})^2\rangle$ of the displacement $z_{i,j}$ within the $j$th layer, and $\Delta z$ is the root mean square of the variance for $j=1,2,\cdots, L_y$: $\Delta z=\sqrt{\sum_j \langle (\Delta z_{i,j})^2\rangle/L_y}$. $\Delta z$ is fluctuating by the noise effect, but sometimes jumps to a fairly large value and then decays stepwise. Figure 10(c) is the time evolution of the kinetic energy $E$. The kinetic energy $E$ jumps when $\Delta z$ jumps. When $\Delta z$ and $E$ jump, phase slips occur between some neighboring layers, which will be shown in Fig.~11(a). Figure 10(d) shows the probability distribution of $v_j$ at $j=20$ in the semi-logarithmic scale. The dashed line is the Gaussian approximation $1/\sqrt{2\pi \sigma^2}\exp\{-(v-0.00345)^2/(2\sigma^2)\}$ with $\sigma^2=0.05025$. The Gaussian approximation is good near $v=0$, however, the deviation from the Gaussian distribution is observed for large $|v|$. The acceleration of $v_j$ at the jumps of $E$ might be related with the velocity distribution with the long tail. 

Nontrivial spatio-temporal dynamics is observed when $\Delta z$ and $E$ jumps.  Figure 11(a) shows the profiles of $z_{i,j}$ as a function of $j$ for $i=25$ at $t=12000+25n$ ($n=0,1,\cdots,30$), which corresponds to the second peak region of $\Delta z$ shown in Fig.~10(b). For example, the lowest line represents a profile of the displacement $z_{i,j}$ at $i=25$ and $t=12000$, and the second lowest one represents $z_{i,j}$ at $i=25$ and $t=12025$. There is a discontinuity at $j=25$ at the profiles of $z_{i,j}$ for $t\le 12375$, which was generated by the phase slip event near $t=5900$ which corresponds to the first peak in Fig.~10(b). Figure 11(a) shows that a phase slip occurs and the discontinuity of $z$ is created at $j=9$ near $t=12400$. And then phase slips occur at $j=34$, and 38 near $t=12800$. The jump of $\Delta z$ near $t=12000$ shown in Fig.~10(b) is caused by these successive phase slips. 
The spatial variation $\Delta z$ of the displacement $z_{i,j}$ decreases stepwise after the jump as shown in Fig.~10(b). The stepwise decrease of $\Delta z$ is understood through the spatio-temporal dynamics of $z_{i,j}$ as follows.  
Figure 11(b) shows the time evolution of $z_{i,j}$ at six points $(i,j)=(1,36)$, $(51,36)$, $(101,36)$, $(151,36)$, $(201,36)$, and $(300,36)$ in the same layer of $j=36$ from $t=13000$ to 15000. It is observed that $z_{i,j}$'s tend to be aligned  with time. The stepwise merging of $z_{i,j}$ makes the spatial variation of $z_{i,j}$ decrease in time, which corresponds to the stepwise decrease of $\Delta z$ shown in Fig.~10(b). Figure 11(c) shows the time evolution of $z_{i,j}$ at the same layer $j=36$. The lowest line is the profile $z_{i,j}$ as a function of $i$ at $j=36$ and $t=14500$, and the second lowest one shows the profile of $z_{i,j}$ at $j=36$ and $t=14525$. There are vortex and antivortex solutions in Eq.~(1) because of the sinusoidal coupling in the $y$ direction. The kink and antikink structures seen in Fig.~11(c) represent the vortex or antivortex.  
Figure 11(c) expresses that  $z_{i,j}$ tends to be uniform through the collisions of vortex and antivortex, which corresponds to the stepwise merging  of $z_{i,j}$ shown in Fig.~11(b) and the stepwise decrease of $\Delta z$ shown in Fig.~10(b). 

The intermittent time evolution of the coupled oscillators at $v_0=0.01$ and $T=0.05$ as shown in Figs.~10 and 11  can be qualitatively explained as follows. The small perturbations $\delta z_{i,j}$ in Eq.~(1) obeys      
\begin{equation}
\frac{d^2 \delta z_{i,j}}{dt^2}=K_x\sum_{i^{\prime}=i+1,i-1}(\delta z_{i^{\prime},j}-\delta z_{i,j})+K_y\sum_{j^{\prime}=j+1,j-1}\cos (z_{i,j^{\prime}}-z_{i,j})(\delta z_{i,j^{\prime}}-\delta z_{i,j})-d\frac{d\delta z_{i,j}}{dt}.
\end{equation}
If the difference $z_{i,j+1}-z_{i,j}$ between the neighboring layers increases with time and is larger than $\pi/2$, $\cos(z_{i,j+1}-z_{i,j})$ becomes negative and the perturbations grow exponentially owing to the linear instability, which induces the phase slips. If the phase slips occur, the kinetic energy jumps and the spatial variation $\Delta z$ in the $x$ direction grows. After the phase slip, the spatial fluctuations tend to decay through the collisions of vortex and antivortex. After $\Delta z$ and $E$ decay to the thermal fluctuation level at the finite temperature $T=0.05$, the differences $z_{i,j+1}-z_{i,j}$ between the neighboring layers increase further again, and phase slips occur at points $(i,j)$ where $\cos(z_{i,j+1}-z_{i,j})<0$, which are generally different from the previous phase-slip points. This process repeats many times. These intermittent time evolutions are observed at small values of $v_0$ and $T$. If the temperature or the shear rate $v_0/L_y$ is large, the spatial irregularity owing to the temperature or deterministic chaos increases and the intermittent behavior of $E$ becomes unclear. 

\begin{figure}[h]
\begin{center}
\includegraphics[height=4.5cm]{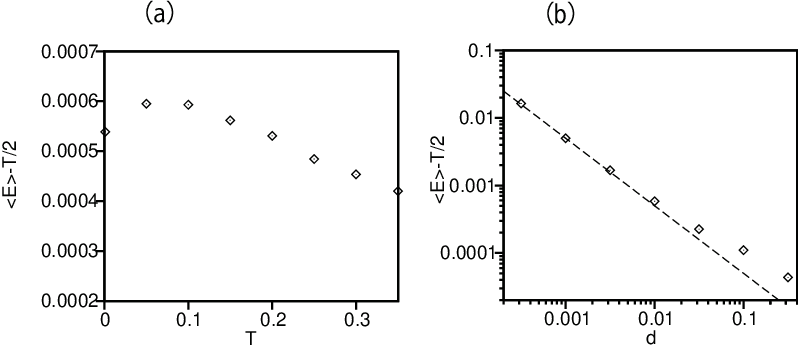}
\end{center}
\caption{(a) Excess kinetic energy $\langle E\rangle -T/2$ for several $T$'s at $d=0.01$ and $v_0=0.01$.  (b) Double logarithmic plot of $\langle E\rangle -T/2$ as a function of $d$ at $T=0.01$ and $v_0=0.01$.}
\label{fig12}
\end{figure}
The three parameters $v_0$, $T$, and $d$ determine the nonequilibrium dynamics in the coupled noisy oscillators Eq.~(5). Although the whole parameter range is not yet investigated and the nonequilibrium dynamics is not theoretically understood, we investigate numerically the excess kinetic energy characterizing the nonequilibrium state by changing some control parameters. The kinetic energy jumps intermittently from the thermal equilibrium level $T/2$ as shown in Fig.~10(c). The excess kinetic energy is the temporal average of the difference of the kinetic energy from $T/2$: $\langle E\rangle -T/2$. Figure 12(a) shows $\langle E\rangle -T/2$ for several $T$'s at $d=0.01$ and $v_0=0.01$. The excess kinetic energy weakly decreases with $T$ for $T>0.1$. Figure 12(b) shows  $\langle E\rangle-T/2$ for several $d$'s in the double-logarithmic scale at $v_0=0.01$ and $T=0.01$. The dashed line is $0.00005/d$. The excess kinetic energy decreases as $1/d$ for small $d$. 

\section{Summary}
We have studied coupled oscillators with inertia under the shear velocity boundary conditions on the rectangular lattice. We have found that the kinetic energy increases with time when $d=0$ or the dissipation is absent. We have calculated the probability distribution of the velocity. When $d=0$ or sufficiently small, the velocity distribution is well approximated at the Gaussian distribution or the Maxwell distribution. We have observed the deviation from the Gaussian distribution when $d$ is large. The deviation from the Gaussian distribution is characteristic of the nonequilibrium state far from equilibrium. We have found that the velocity profile changes from the linear one to the localized one near the boundary as $d$ is larger. In the coupled oscillators under a finite temperature, the localization becomes weaker, and the probability distribution of the velocity becomes closer to the Gaussian distribution. When $v_0$ is small and $T$ is small but not zero, the coupled oscillators are a weakly nonequilibrium stochastic system. We have observed intermittent jumps of the kinetic energy. The jumps correspond to the phase slips in the $y$ direction, and an annealing process of spatial fluctuations of the displacement $z_{i,j}$ in the $x$ direction occurs through the vortex-antivortex collisions. 

Our coupled oscillator model under the shear velocity boundary condition is an interesting system, since the model exhibits various nonlinear-nonequilibrium phenomena and is related to various research areas such as the coupled limit-cycle oscillators, the dislocation theory in solids, a block-spring model of earthquakes, and the nonequilibrium molecular dynamics. We have shown several numerical results in this paper, however, the theoretical understanding is not sufficient, which is left to future study. 


\begin{thebibliography}{99}
\bibitem{Kuramoto} Y.~Kuramoto, {\it Chemical Oscillations, Waves, and Turbulence} (Springer, New York, 1984). 
\bibitem{Bonilla} J.~A.~Acebr\'on, L.~L.~Bonilla, C.~J.~Perez Vincente, F.~Ritort, and R.~Spigler, Rev. Mod. Phys. {\bf 77}, 137 (2005). 
\bibitem{Strogatz} S.~Strogatz, Physica D {\bf 143}, 1 (2000). 
\bibitem{Tanaka} H.~Tanaka, A.~J.~Lichtenberg, and S.~Oishi, Phys. Rev. Lett. {\bf 78}, 2104 (1997).
\bibitem{Sakaguchi} H.~Sakaguchi and T.~Matsuo, J. Phys. Soc. Jpn. {\bf 81}, 074005 (2012).
\bibitem{Sakaguchi2} H.~Sakaguchi, S.~Shinomoto, and Y.~Kuramoto, Prog. Theor. Phys. {\bf 77}, 1005 (1987).
\bibitem{Hong} H.~Hong, H.~Chate, H.~Park, and L.~H.~Tang, Phys. Rev. Lett. {\bf 99}, 184101 (2007).
\bibitem{Sakaguchi3} H.~Sakaguchi, Phys. Rev. E {\bf 105}. 054211 (2022).
\bibitem{Read} W.~T.~Read, {\it Dislocations in crystals} (McGraw Hill, 1953).
\bibitem{Cottrell} A.~H.~Cottrell {\it Dislocations and plastic flow in crystals} (Clarendon Press, 1953).
\bibitem{Sakaguchi4} H.~Sakaguchi, Phys. Rev. E {\bf 106}, 054154 (2022).
\bibitem{Scholz} C.~H.~Scholz, {\it The Mechanics of Earthquakes and Faulting} (Cambridge University Press, Cambridge, 2002).
\bibitem{Langer} J.~M.~Carlson and J.~S.~Langer, Phys. Rev. Lett. {\bf 62}, 2632 (1989).
\bibitem{Sakaguchi5} H.~Sakaguchi and S.~Kadowaki, J. Phys. Soc. Jpn. {\bf 86}, 074001 (2017).
\bibitem{Evans} D.~J.~Evans and G.~P.~Morriss, {\it Statistical Mechanics of Nonequilibrium Liquids} (Academic Press, San Diego, 1990).
\bibitem{Ashurst} W.~T.~Ashurst and W.~G.~Hoover, Phys. Rev. A {\bf 11}, 658 (1975). 
\bibitem{Liem} S.~Y.~Liem, D.~Brown, and J.~H.~R.~Clarke, Phys. Rev A {\bf 45}, 3706 (1992).
\bibitem{Lepri} S.~Lepri, R.~Livi, and A.~Politi, Phys. Rev. Lett. {\bf 78}, 1896 (1997).
\bibitem{Holian} B.~L.~Holian, W.~G.~Hoover, and H.~A.~Posch, Phys. Rev. Lett. {\bf 59}, 10 (1987).
\bibitem{Bingham} E.~C.~Bingham, {\it Fluidity and Plasticity} (MacGraw-Hill, New York, 1922).
\bibitem{Salmon} J.~B.~Salmon, A.~Colin, S.~Manneville, and F.~Molino, Phys, Rev. Lett. {\bf 90}, 228303 (2003). 
\bibitem{Rutter} E.~H.~Rutter, Techtonophysics {\bf 303}, 147 (1999).
 
\end{thebibliography}
\end{document}